\documentstyle[prl,aps]{revtex}
\begin{document}
%------------------------------------------------------------------------------
\title{Parametrization of Multiple Pathways in Proteins:
Fast Folding versus Tight Transitions}
\author{Paul G.\ Dommersnes,\footnote{Permanent address: Department of Physics,
Norwegian University of Science and Technology, NTNU, N--7491 Trondheim,
Norway} Alex Hansen,$^*$ Mogens H.\ Jensen, and Kim Sneppen}
\address{Niels Bohr Institute and NORDITA, Blegdamsvej 17, DK--2100 
Copenhagen {\O}, Denmark}
\date{\today}
\maketitle
%------------------------------------------------------------------------------
\begin{abstract} 
Growing experimental evidence shows that proteins follow one or a few
distinct paths when folding.
We propose in this paper a procedure to 
parametrize these observed pathways, and from this parametrization 
construct effective Hamiltonians for the proteins. We furthermore study the
denaturated-native transitions for a wide class of possible effective 
Hamiltonians based on this scheme, and find that the sharpness (tightness) of
the transitions typically are close to their theoretical maximum
and thus in quantitative accordance with the sharp folding transition
observed for single domain proteins.
Finally we demonstrate that realistic folding times are typical
for the proposed class of Hamiltonians, and we discuss the 
implication of the predicted entropy barriers on the 
temperature dependence of the folding times.
\end{abstract} 
%------------------------------------------------------------------------------
\bigskip
%------------------------------------------------------------------------------
\section{Introduction}
\label{intro}
%------------------------------------------------------------------------------
Globular proteins exists in two states 
the native, or folded, state and the denaturated,
or unfolded, state \cite{c93}. 
The folding of a protein is mostly driven by thermodynamics,
and it typically occurs on a time scale of order seconds. 
The folding, and the resulting conformations result from an interplay 
between polar
and nonpolar amino acids on the protein, their hydrogen bonds,
and how all these interact with and constrain the surrounding water.
The folding is therefore an immensely
complex process and to date it is still a completely open problem how
one may predict a protein's folding properties given its structure.

On a qualitative level it is remarkable that the 
native state of a globular protein is solid with specified
positions of all amino acids. This fact is thermodynamically 
reflected in a denaturation phase transition which is exceedingly sharp, 
in spite of the small number of molecules in the protein.
Enhancing this qualitatively remarkable phenomenon is the fact that
a typical polymer folds with a very weak transition \cite{s80}.
In fact single domain protein folding transition fairly accurately
behaves as a two-state thermodynamical system \cite{p96} where only
denatured and native state can be populated and 
nearly no states with energies
between these two extremal states are observed.
Another remarkable feature of proteins is that they reach their unique ground
state in a time that is much smaller than the time needed to
search the extremely large configurational space.
This is commonly known as the Levinthal paradox,  and is noteworthy
because it is difficult to merge with the two-state thermodynamics
of proteins. 

Research into the more specific aspects of the dynamics of protein
folding is divided into several sub camps.  For example, there is a
huge effort to study simplified models numerically using ideas
originating from spin glass theory. The basic idea here is that the
different possible conformations of a protein define an energy
landscape containing multiple minima. Such a picture is supported by
experimental evidence at very low temperatures and structures confined 
by potentials less than about 2kcal/mol \cite{f97}.  
However, if the dynamics of proteins at normal
temperatures were that of spin glasses, its folding would be even slower
than a random search in configuration space, 
resulting in time scales on an astronomical scale.  This is not the case.
Today, one imagines two possible ways to explain how proteins 
can fold on reasonable timescales: (1) The 
existence of {\it folding funnels\/} \cite{bosw95,d95,d99}, or (2)
the existence of {\it folding pathways\/} \cite{br99a,br99b}.
The basic idea behind the
folding funnel is that it is a deformation of the energy landscape in
such a way that it resembles a funnel.  The protein then ``falls" into
the global ground state without getting stuck. 

The folding pathway is a concept related to a specific sequence of events.  
In contrast to a funnel, a pathway is characterized by subsequent formation
of identifiable structures that must be formed in a specific order.
Pathway studies of proteins that fold have been extensively carried 
out by Fersht \cite{f93}. Occasionally one observes more than 
one possible pathway. The mechanism leading to the
existence of guiding pathways may be 
hierarchical from small scales to larger scales in the sense of 
Baldwin and Rose \cite{br99a,br99b}, 
or it may be purely sequential as suggested in \cite{hjsz98b}.
It is the aim of this paper to demonstrate a third possibility, that
encompass both of the above suggested scenarios but do it in a way
that ensure the experimentally observed two state nature 
of protein folding-unfolding.

In the next section, we introduce and discuss our parametrization scheme for
protein folding.  This is based on our earlier work on one-pathway protein
folding \cite{hjsz98b,hjsz98a,hjsz99}.  
In order for a protein to be biologically
relevant, certain demands has to be put on its folding properties.  One such
demand is that the folding-unfolding transition is sharp.  The sharpest possible
transition is obtained when the system is two state.  We analyse in
Section \ref{sharpness} a class of Hamiltonians parametrized by the scheme
outlined in Section \ref{multipaths} with respect to sharpness using the
van't Hoff relation.  We find that the {\it typical\/} system within this class
behaves as a two-state system with maximal sharpness of transition.
In Section \ref{time}, we finally demonstrate that the class 
of Hierarchical folders encoded by the parametrization scheme of
Section \ref{path} have folding times compatible with the ones that
are experimentally observed. 
Further we demonstrate that the phase transition is sharpened 
by an entropic barrier and therefore that the folding time increase strongly
with temperature $T$ in some $T$ interval below the equilibrium transition
temperature $T_c$. 
In Section \ref{conc} we conclude that both demands, 
the sharpness of transition and the short folding times, 
are encoded by sequentially encoded hierarchical folders. 
%------------------------------------------------------------------------------
\section{Pathways}
\label{path}
%------------------------------------------------------------------------------
We take the point of view that proteins follow certain 
pathways when folding from
the denaturated to the native state.  The ``standard" theorist's way 
of envisionin such a pathway is as a curve in some rugged energy landscape
leading to the global minimum point --- which corresponds to the native
state.  In the present work, we take a different point of view.
We define a pathway as a {\it sequence of partially folded structures\/} 
which sequentially reduces both energy and entropy of the protein,
and thus guide it towards the native state.
By a partially folded structure 
we mean a protein conformation where parts of it 
has taken on the structure of the native state.  

The simplest model of subsequent entropy pruning is a zipper \cite{dfc93}, 
where one at each step $k$
of the folding gains one unit of energy, say $E(n+1)=E(n)-a$, and 
simultaneously reduces the available number of conformational states 
with a factor $g$,
i.e., state space $S(k) \rightarrow S(k+1) \;=\; S(k)/g$.
If each step is associated with binding of one amino acid from a random coil
then $g\sim 6$ corresponding to the possible orientations of one amino acid
along the chain \cite{c93}. 
If we deal with reduction of effective number of orientations
form a compact molten globule, then $g \sim  6/e \sim 2$ where the factor $e$
comes from excluded volume effects \cite{g93}. For individual amino acid
contacts the corresponding energy
gains may be estimated from the Miyazawa-Jernigan matrix \cite{mj85}, 
suggesting energies $a$ of order $1 \rightarrow 2$ kcal/mol.
In the first subsection we will explore the mathematical formulation of this
simple zipper model, and demonstrate that it indeed predicts a 
first order phase transition, albeit a fairly weak one.

Folding of real proteins are presumably more complicated than a simple zipper.
Protein structure data collected by Baldwin and Rose \cite{br99a,br99b} 
support that the native state of real proteins is reached from progressively 
larger parts of the protein which are stabilized by the smaller structures 
already in place. This indeed suggest a sequential ordering of events with some similarity
to the zipper, but with the important difference that folding may initiate 
at different places along the amino acid chain simultaneously. 
If this was the case, the resulting independence of events would imply
a factorization of the partition function into sub-parts, with the result
of weakening or eventually destroying the phase transition altogether.
As discussed in Section \ref{sharpness}, 
the melting phase transition of a single domain protein is remarkably sharp, 
implying that folding the protein is a very cooperative and correlated process.

We will in Section \ref{sharpness} see that the observed sharpness 
of the phase transition of proteins puts severe constraint
on any protein folding model, e.g.\  ruling out both the simplest version
of the zipper model and the simplest version of the hierarchy model.
In addition, there is growing evidence that proteins 
may follow alternate pathways during the folding process 
\cite{p96,br99b,hrssm99,cp00,lgrgd00}, thus requiring
a formalism that merge the conflicting interest between the need
of dealing with more than one pathway, to fold in a reasonable time, and to
maintain the native state sharply folded when it finally is reached.

The purpose of this section is to develop such a formalism, thus encompassing
both the pathway formalism and the possibility that folding proceed 
hierarchically from smaller to larger structures.
We first assume that there is only one pathway. 
This will allow us to introduce the
central ideas in our approach in the next subsection. In subsection 
\ref{multipaths} we subsequently generalize theses ideas into a  
scenario where several pathways are allowed.   
%------------------------------------------------------------------------------
\subsection{Single Pathway}
\label{singlepath}
%------------------------------------------------------------------------------
A folding pathway is a sequence of partially folded structures. Each
structure along the pathway may contain the previously folded
structures as a subset.  If there is only one pathway available for
the protein, this will be the case.  We are therefore dealing with a
hierarchy where structures use previously folded structures as
building blocks.  It is also clear that the folding process towards
the native state proceeds from smaller to larger scales.

We number the partially folded structures as they appear in the
folding process, $i=1,\cdots,N$.  The binary variable $\psi_i$ equals
1 if partially folded structure $i$ exists --- also if it forms part
of an even larger folded structure.  Otherwise, it is zero. Thus, the
folding pathway may be parametrized by the sequence of inequalities
\begin{equation}
\label{eq:singlepath}
\psi_1\ge\psi_2\ge\psi_3\ge\cdots\ge\psi_N\;.
\end{equation}
We will in the following refer to the $\psi_i$ variables as {\it
folding variables.\/}

There is energy and entropy associated with the partially folded
structures.  There are the direct binding energies that keeps the
folded structures together.  Perturbing the folded parts of the
protein requires more energy than perturbing the unfolded parts.  The
latter may change conformation with very low or no energy cost.  Thus,
these changes appear as entropy on the thermodynamical level.  In
order to model these mechanisms, we introduce a second set of
variables, the {\it structural variables\/} $\xi_1$ to $\xi_N$.  These
variables contain e.g.\ the angles $\phi$ and $\psi$ associated with
$N-C_\alpha$ and the $C_\alpha-C'$ bonds of the part of the protein
that folded.  However, they are simplifications --- and to simplify
even further, we assume that they only take two values \cite{hjsz99},
\begin{equation}
\label{eq:xiento}
\xi_i=\cases{1\;,\cr
             1-\Xi\;.\cr}
\end{equation}
This latter simplification is not essential to retain an analytically
tractable model.  However, it makes it easier to introduce the model and
discuss it in general terms.
The choice of two states corresponds to a degeneracy factor $g=2$, and
one can easily generalize the results we will obtain to larger $g$
values by simply assuming that $g-1$ states are associated with the
higher energy level.

We now proceed to construct the simplest possible Hamiltonian that
reflect the above discussion \cite{hjsz99}.  It is
\begin{equation}
\label{eq:singpatham}
{\cal H}=-\sum_{i=1}^N a^{(i)}\psi_i\xi_i\;,
\end{equation}
where $a^{(i)}$ are positive coupling constants.  

When the protein is at stage $j$ in the folding process, the energy associated
with it is
\begin{equation}
\label{eq:stagejenergy}
E_i=\sum_{i=1}^j \xi_i\;.
\end{equation}
The parts of the protein not yet folded provide entropy, which in this 
particular case consists of the variables $\xi_{j+1},\cdots,\xi_N$.
The degeneracy of this protein configuration is $2^{N-j}$.

Working with the Hamiltonian Eq.\ (\ref{eq:singpatham}) with the constraints
(\ref{eq:singlepath}) is somewhat cumbersome.  We will therefore
make a coordinate transformation that builds the constraints into the
Hamiltonian directly.  As we will see in the next section, this transformation
allows us to generalize the one-pathway Hamiltonian presented in this section to
multiple-pathway Hamiltonians:  Specify a set of pathways, and the 
corresponding Hamiltonian can be written down immediately.

We assume in the following that the value $\Xi$ defined in Eq.\
(\ref{eq:xiento}) is so large that the $\xi_i$'s for which $\psi_i=1$,
will not take on this value for any realistic temperature ($T<100$ C).
In this case, we may set
\begin{eqnarray}
\label{eq:transf}
\psi_1&=&\phi_1\;,\nonumber\\
&\vdots&\nonumber\\
\psi_i&=&\phi_1\phi_2\cdots\phi_i\;,\nonumber\\
&\vdots&\nonumber\\
\psi_N&=&\phi_1\phi_2\cdots\phi_i\cdots\phi_N\;,\nonumber\\
\end{eqnarray}
where there are no constraints on the new variables $\phi_i$.  We see
that the constraints Eq.\ (\ref{eq:singlepath}) automatically are obeyed.
The Hamiltonian (\ref{eq:singpatham}) now becomes
\begin{equation}
\label{eq:singpathphi}
{\cal H}=-a^{(1)}\phi_1-a^{(2)}\phi_1\phi_2-\cdots-a^{(N)}\phi_1\cdots\phi_N\;,
\end{equation}
We furthermore note that the degeneracy provided by the $\xi_i$
variables (as they may take on two possible values, 1 and $1-\Xi$,
when the corresponding $\psi_i$ is zero), is also build into the
Hamiltonian.  Assume e.g.\ that all $\phi_1$ to $\phi_j$ are equal to
one,  this corresponds to $\psi_1=\cdots\psi_j=1$.  Furthermore
assume that $\phi_{j+1}=\psi_{j+1}=0$. The remaining $\phi_{j+2}$ to
$\phi_N$ can take on any value without change in energy.  Thus, to
within a factor 2, the degeneracy is the same that provided by the
$\xi_i$ variables in the original Hamiltonian (\ref{eq:singpatham}).

The Hamiltonian (\ref{eq:singpathphi}) has a first order transition at
a temperature $T=1/\log2$ when $a^{(1)}=\cdots=a^{(N)}=1$
\cite{hjsz98b}. This Hamiltonian is easily generalized to take into
account the coupling between water and protein \cite{hjsz99}.
%------------------------------------------------------------------------------
\subsection{Multiple Pathways}
\label{multipaths}
%------------------------------------------------------------------------------
We need to define what we mean by ``multiple pathways."  We return to the
single pathway defined using the folding variables 
by Eq.\ (\ref{singlepath}).  Suppose we break this
sequence of inequalities by removing one of them, say $\psi_i\ge\psi_{i+1}$,
so that we have
\begin{equation}
\label{eq:twopaths1}
\psi_1\ge\cdots\ge\psi_i\;,
\end{equation}
and
\begin{equation}
\label{eq:twopaths2}
\psi_{i+1}\ge\cdots\ge\psi_N\;.
\end{equation}
Now, the status (0 or 1) of any $\psi_j$, where $j\le i$ is independent of
the status of any $\psi_k$, where $k >i$.  That is, we have created two
independent folding ``domains."   

In terms of the $\phi_i$ variables, the two groups of inequalities 
(\ref{eq:twopaths1}) and (\ref{eq:twopaths2}), may be expressed as
\begin{eqnarray}
\label{eq:twopathsphi1}
\psi_1&=&\phi_1\;,\nonumber\\
&\vdots&\nonumber\\
\psi_i&=&\phi_1\cdots\phi_i\;,\nonumber\\
\end{eqnarray}
and
\begin{eqnarray}
\label{eq:twopathsphi2}
\psi_{i+1}&=&\phi_{i+1}\;,\nonumber\\
&\vdots&\nonumber\\
\psi_N&=&\phi_{i+1}\cdots\phi_N\;.\nonumber\\
\end{eqnarray}
We may also construct the corresponding Hamiltonian,
\begin{equation}
\label{eq:hamham}
{\cal H}=-a^{(1)}\phi_1-a^{(2)}\phi_1\phi_2-\cdots-a^{(i)}\phi_1\cdots\phi_i
-a^{(i+1)}\phi_{i+1}-\cdots-a^{(N)}\phi_{i+1}\cdots\phi_{N}\;.
\end{equation}

When setting up the Hamiltonian (\ref{eq:hamham}), we have made the assumption
that both pathways, $1\to 2\to\cdots\to i$ and $(i+1)\to\cdots\to N$ have the
same degeneracies associated with them, fixing the relative entropic 
contribution from each branch.  By introducing additional independent
variables $\phi_{N+j}$, we may tailor these contributions.  For example, 
the Hamiltonian 
\begin{equation}
\label{eq:hamham1}
{\cal H}=-a^{(1)}\phi_1-a^{(2)}\phi_1\phi_2-\cdots-a^{(i)}\phi_1\cdots\phi_i
-a^{(i+1)}\phi_{i+1}\phi_{N+1}-\cdots-a^{(N)}\phi_{i+1}
\cdots\phi_{N}\phi_{N+1}
\end{equation}
will increase the entropic contribution of the second branch by 
$T\log 2$ compared to the first branch.

We may generalize the ideas presented above to systems with multiple folding 
pathways.  This is done simply by generating telescoping groups of $\phi_i$ 
variables, one for each pathway.  The corresponding Hamiltonian may be
written down directly, 
\begin{equation}
\label{eq:calhissum}
{\cal H}=-\sum_{i}a^{(1)}_i\phi_i-\sum_{i \ne
j}a^{(2)}_{ij}\phi_i\phi_j-\dots- \sum_{i \ne j \ne k
\dots}a^{(N)}_{ijk \dots}\phi_i\phi_j\phi_k\dots\;.
\end{equation}
We have here added subscripts to the $a^{(i)}$ coefficients, reflecting which
$\phi_i$ variables are involved in the corresponding term in the Hamiltonian. 

In order to construct a Hamiltonian given a set of pathways, we start
with organizing the pathways in a branching structure where each
numbering refer to a folded substructure of the protein, as shown in
Fig.\ \ref{fig1}.  The interpretation of this figure is that folding
of structures 1 and 5 are independent starting points of the folding
process.  In order to structure 4 to be formed, structure 5 must
already be in place.  However, this structure is independent of
structure 1 having been formed.  Structure 2, on the other hand, needs
both structures 1 and 5 in place.  Lastly, structure 3 can only form
if both structures 2 and 4 have formed. In the end, all structures
must always be formed. A concrete example of such a set of pathways has been
reported for staphylococcal nuclease \cite{ws95,swgw96}.

Assuming now that the reduction in degeneracy when either structure 1 or 5
form is the same, we may set up a ``folding table" as shown in Table
\ref{table1}.  From this table, we read off the Hamiltonian
\begin{equation}
\label{eq:hamhamex}
{\cal H}=-a^{(1)}_1\phi_1-a^{(1)}_5\phi_5-a^{(2)}_{45}\phi_4\phi_5
-a^{(3)}_{125}\phi_1\phi_2\phi_5-a^{(5)}_{12345}
\phi_1\phi_2\phi_3\phi_4\phi_5\;.
\end{equation}

In the following, we will study the thermodynamical and dynamical
properties of Hamiltonians constructed in this way.  There are two
main questions to be addressed which are essential in connection with
protein folding: (1) sharpness of folding transition and (2) shortness
of folding time.  In the next section, we will consider the sharpness
of the transition.
%------------------------------------------------------------------------------
\section{Sharpness of Transition}
\label{sharpness}
%------------------------------------------------------------------------------
An excellent measure of sharpness of a first-order transition is to study the
van't Hoff relation \cite{pk74,p97} which relates enthalpy change through the
transition $\Delta H$, the peak of the heat capacity $C_{\rm p}$, the 
absorbed heat, $Q$ through the formula
\begin{equation}
\label{eq:vanthoff}
\Delta H = \alpha\, \frac{T_{\rm c}^2 C_{\rm p}( T_{\rm c}) }{Q}\;.
\end{equation}
Here $T_{\rm c}$ is the transition temperature, and $\alpha$ is the 
{\it van't Hoff coefficient.\/}  The smaller this coefficient is, the sharper
the transition.

We may demonstrate this in the following way.
In the present case, $\Delta H = Q$ (as there is no pressure in the system),
which is the area of the bump in the
heat capacity plot.  The height of the hump, $h$, is $C_{\rm p}$, 
while the area $Q$ is essentially the width of the top $b$ multiplied by the 
height, $h$, $Q=bh$.  We may thus write the van't Hoff equation
\begin{equation}
\label{eq:vtt}
b^2 h = \alpha T_{\rm c}^2\;.
\end{equation}
>From this equation it might seem that the opposite of what we are
claiming is true: The larger the $\alpha$, the wider the transition,
as the height $h$ is proportional to $\alpha$.  However, one must not
forget the presence of the width $b$.  We rewrite the left hand side
of Eq.\ (\ref{eq:vtt}) as
\begin{equation}
\label{eq:vtt1}
(bh)^2\ {1\over h} = \alpha T_{\rm c}^2\;.
\end{equation}
Now keeping the area of the hump, $bh=Q$, constant, we see that the height
$h$ is {\it inversely\/} proportional to $\alpha$.  Hence, our statement
that a smaller $\alpha$ implies a sharper transition follows, as sharpness
is a relative concept.

In order to understand the the $\alpha$ coefficient properly, we
examine its physical meaning in the following.  Let us assume a system with 
an equally spaced energy spectrum, 1,...,$N$.  We furthermore assume a
degeneracy $g_n$ for the $n$th energy level.  The partition function is then
\begin{equation}
     \label{eq:partz}
   Z=\sum_{n=0}^N g_n {\rm e}^{n/T}   \;.
\end{equation}
For proteins there should be a unique ground state, so that $g_N=1$. 
The energy is given by
\begin{equation}
   \label{eq:energy}
   E(T)=T^2\frac{\partial}{\partial T} \log{Z}=-\langle n \rangle\;,
\end{equation}
and the heat capacity is given by 
\begin{equation}
\label{eq:heatcapacity}
 C(T)=\frac{\partial}{\partial T} E(T)=\frac{\langle n^2
 \rangle-{\langle n \rangle}^2}{T^2}\;.
\end{equation}
These relations are of course valid at any temperature. The $\alpha$ factor
can now be expressed in terms of the fluctuations of the energy levels
at $T=T_{\rm c}$,
\begin{equation}
\label{eq:fluctnn}
 \langle n^2 \rangle-{\langle n \rangle}^2=\frac{N^2}{\alpha}\;.
\end{equation}
For {\it any} probability distribution we must have $\langle n^2
\rangle-{\langle n \rangle}^2 \le N^2/4$. We only have equality when
$P(n=0)=P(n=N)=1/2$. Thus in general 
\begin{equation}
\label{eq:alpha4}
\alpha \ge 4\;. 
\end{equation}
Equality $\alpha=4$ is only found when the system
exhibits two state behaviour, i.e., when $g_0=g_N{\rm e}^{N/T}$ 
and otherwise $g_n=0$.  This is fulfilled with the Hamiltonian 
${\cal H}=-N \phi_1 \cdots \phi_N$, i.e., when all 
terms $a^{(i)}=0$ for all $i < N$ in Eq.\ (\ref{eq:hamham}).
In general Eq.\ (\ref{eq:fluctnn}) can be reformulated as
\begin{equation}
\label{eq:fluctnn1}
 \langle s^2 \rangle-{\langle s \rangle}^2=\frac{1}{\alpha}\;.
\end{equation}
which express $\alpha$ as the variance of the distribution of
normalized energy states ($s=n/N$) at the transition point.  This
distribution could be bimodal, ($\alpha=4$), be flat ($\alpha=12$) or
be centered around a mean intermediate energy ($\alpha > 12$).  The
flat distribution is obtained when all $a^{(i)}=const$ in the
Hamiltonian (\ref{eq:hamham}).  Of special interest is the class of
Hamiltonians that correspond to the hierarchical folding suggested by
Baldwin and Rose \cite{br99a,br99b}.  We show in Fig.\ \ref{fig1.5} an
example of such a hierarchical folding network with four levels.  In
this scheme the folding of smaller units form a necessary but not
sufficient condition for the folding of larger units.  A corresponding
Hamiltonian may be constructed as outlined in Section
\ref{multipaths}, and we find
\begin{equation}
\label{eq:baldrose}
{\cal H}\; = \; - a^{(1)}_1\phi_{1}- a^{(1)}_2\phi_{2}- a^{(1)}_3\phi_{3} - 
a^{(1)}_4\phi_{4}
- a^{(3)}_{125}\phi_{1} \phi_{2} \phi_5 -  a^{(3)}_{346}\phi_{3} \phi_{4}\phi_6
- a^{(7)}_{1234567}\phi_{1} \phi_{2} \phi_{3} \phi_{4}\phi_5\phi_6\phi_7\;.
\end{equation}
At each level in the hierarchy a new variable is needed in order to make 
folding at level below a necessary but sufficient condition for folding at
this level.

As the number of terms in this hierarchical Hamiltonian grows, $\alpha
\rightarrow \infty$.  The reason for this is the statistical
broadening due to the exponentially growing number of independent
branches when moving ``backwards" in the hierarchy towards smaller
units.

Privalov and Khechinasvili \cite{pk74} measured experimentally the value
\begin{equation}
\label{eq:4.2}
\alpha \; =\; 4.2 \pm 0.2\;.
\end{equation}
for large group of small globular proteins.  Thus real single domain
proteins are close to two-state behaviour, i.e., far from a simply
guided Hamiltonian, and very far from protein folding viewed as
subsequent folding of independent subunits.  Folding of single domain
proteins, that in practical terms means proteins of about 100 amino
acids is indeed close to a maximally cooperative process.
%------------------------------------------------------------------------------
\subsection{A Class of Hierarchical Hamiltonians with $\alpha=4$.}
\label{realmodel}
%------------------------------------------------------------------------------
Based on spatial correlations of amino acids in real proteins, Baldwin
and Rose argue convincingly that some sort of subsequent folding
hierarchy must be implemented in real protein folding
\cite{br99a,br99b}.  In last section we saw that the simplest
hierarchical scheme is excluded for single domain proteins. We now
explore a more elaborate hierarchical scheme, where correlations
between folding subunits naturally becomes implemented on an early
level.  The simplest of these is the single pathway model of Eq.\
(\ref{eq:singpathphi}).  A general formulation is the Hamiltonian
(\ref{eq:calhissum}).  It has up to $N$ groups of products of various
length of $\phi_i$ variables.  We will in next subsection investigate
the broad subclass of these Hamiltonians which are restricted by having
maximum one term of each product number, and see that they typically
have $\alpha \approx 4$.  This means that we investigate Hamiltonians
as in Eq.\ (\ref{eq:calhissum}) 
where the coefficient $a^{(m)}_{ijk\dots}=1$ for one
combination of $ijk\dots$ and otherwise zero.  Thus, there is one term
with only one $\phi_i$, one term with two $\phi_i$ variables etc.

We note that the folding system in Fig.\ \ref{fig1} as described by
Eq.\ (\ref{eq:hamhamex}) does not include only one term at each level.
However, the system in Fig.\ \ref{fig1} can also be parametrized by
introducing one additional auxiliary variable $\phi_6$, thereby
creating an entropy barrier in the thermodynamics.  The corresponding
Hamiltonian for Fig.\ \ref{fig1} would then be 
\begin{equation}
\label{eq:hamhamhamex}
{\cal H}=-a^{(1)}_1\phi_1-a^{(1)}_5\phi_5 \phi_6-a^{(3)}_{456}\phi_4
\phi_5\phi_6 -a^{(4)}_{1256}\phi_1\phi_2\phi_5\phi_6-a^{(6)}_{123456}
\phi_1\phi_2\phi_3\phi_4\phi_5\phi_6 \;.
\end{equation}
This is the type of Hamiltonians that we wish to study.  This subclass
deals with systems where indeed different regions may start
independent folding, but where some foldings are more difficult to
initiate and thus becomes rate limiting.  In other words, that
different events takes very different times implies that some folding
events acts as effective nucleation centers.  When these barriers are
overcome, subsequent foldings spread across the system along some
fairly well defined paths.

The nucleation barriers were associated to cases where several variables
must fold simultaneously without previous guiding ($\phi_5$ and
$\phi_6$ in the above Hamiltonian (\ref{eq:hamhamhamex})).  In
addition, different folding paths typically become correlated
e.g.\ through a common origin, as in the above Hamiltonian where the
events $\phi_5$ and $\phi_6$ both are needed in order to proceed
folding along the parallel ways parametrized by respectively $\phi_2$
and $\phi_4$.  Such correlation could e.g.\ be due to steric
interactions that facilitate folding by bringing far-apart amino
acids closer to each other.  These associated entropy barriers and
correlations are responsible for the effective two state behaviour
that this class of hierarchical models exhibit. We will examine this
in detail in the next section.
%------------------------------------------------------------------------------
\subsection{Averaging over the Class of Hamiltonians}
\label{randscheme}
%------------------------------------------------------------------------------
We will in following show that the type of Hamiltonians discussed in
the previous section displays remarkable thermodynamic properties.  We
shall consider more specificly the class of Hamiltonians defined by
Eq.\ (\ref{eq:calhissum}), 
where there is only one term for each number of $\phi_i$s.
The partition function for a Hamiltonians belonging to this class is
\begin{equation}
\label{eq:calz}
Z=\sum_{ \{ \phi \} } {\rm e}^{-{\cal H}/T}\;.
\end{equation}
To gain insight into the thermodynamics of this class of Hamiltonians,
we have calculated the mean of the partition function over all
possible Hamiltonians. The result is
\begin{equation}
\label{eq:z}
 \langle Z \rangle\; = 1+\sum_{s=1}^N \left(
\begin{array}{c} N\\s
\end{array}\right)\prod_{k=1}^{s}\left[1+\frac{s!(N-k)!}{N!(s-k)!}({\rm
e}^{1/T}-1)\right]\;.
\end{equation}
The calculation is outlined in Appendix \ref{ap:average}. This gives
us the thermodynamics of an average Hamiltonian. In order to find out
if this is a representative result we have to calculate the
fluctuations around this mean.  All thermodynamics properties of a
given Hamiltonian is derived from $\log Z$, and the mean can be
expressed as $\langle \log Z \rangle =\langle \log z\rangle
+\log{\langle Z \rangle }$, with $z=Z/\langle Z\rangle$. We have in
Appendix \ref{ap:selfaverage} transformed the expression for the
fluctuations of the partition function into the function, Eq.\
(\ref{eq:amss}) and the result is plotted in Fig.\ \ref{fig2}. The
fluctuations are extremely small, in fact $\langle z^2 \rangle -\langle
z\rangle^2\rightarrow 0$ as the system size increases. For example for
for $N\sim100$ they contribute $\sim10^{-4}k_{\rm B}T$ to the free
energy, and hence without any physical relevance. In other words two
randomly chosen Hamiltonians will have exactly the same thermodynamic
properties, the system is {\it self-averaging\/} and Eq.\ (\ref{eq:z}) is the
exact solution of the partition function. 

In order to study the folding transition, we define an order parameter, $u$,
by the mean of the sum over all the $\phi$ variables 
\begin{equation}
\label{eq:uisvar}
  u=\frac{2}{N}\sum_{i=1}^{N}\langle \phi_i \rangle-1\;.
\end{equation}
When the protein is in the ground state $\langle \phi_i \rangle=1$ and
the order parameter is $u=1$, at high temperatures $\langle \phi_i
\rangle=1/2$ and $u=0$.  Since the class of hamiltonians that we study
are self-averaging for large $N$, we may calculate the thermal
mean values by taking the average over all Hamiltonians. We may thus
write
\begin{eqnarray}
\label{eq:uisfrac1}
 \sum_i\langle \phi_i\rangle &=& \frac{1}{M}\sum_{\{ a \}} \sum_{ \{
\phi \}} \sum_i \phi_i   {\rm e}^{-{\cal H}/T}/\langle Z \rangle
\nonumber \\ &=&\frac{1}{ \langle Z \rangle }\sum_{s=1}^N s\left(
\begin{array}{c} N\\s
\end{array}\right)\prod_{k=1}^{s}\left[1+\frac{s!(N-k)!}{N!(s-k)!}({\rm
e}^{1/T}-1)\right]\;,
\end{eqnarray}
where $M$ is the total number of Hamiltonians.  Fig.\ \ref{fig3} shows
the order parameter at different temperatures. We clearly see that
there is sharp first-order transition around the temperature $T_{\rm
c}\sim 1.44$. The sharpness of the transition is characterized by the
van't Hoff $\alpha$ coefficient, which is calculated from the
partition function Eq.\ (\ref{eq:z}). The results plotted in Fig.\
\ref{fig4} show that it goes quickly to $\alpha=4$ for large $N$,
which is the lowest possible value, i.e.\ the sharpest possible
transition. We conclude hence that the folding scheme we have
introduced gives a first-order phase transition with a van't Hoff
coefficient of $\alpha\sim 4$, corresponding to a {\it two-state\/}
thermodynamics.

We note that for the single-path Hamiltonian, Eq.\
(\ref{eq:singpathphi}) with all $a^{(i)}=1$, $\alpha$ equals 12.  However, if
the last term is set equal to $a^{(N)}=\epsilon N$, where
$\epsilon\to0$ is small (and goes to zero as $N$ in the limit $N \to
\infty $), then $\alpha \approx 4$ \cite{bhhsj00}.
%------------------------------------------------------------------------------
\section{Folding Time}
\label{time}
%------------------------------------------------------------------------------
Proteins in their native form have well defined structures that are
separated from their denatured counterparts through some barrier, as
quantified by value of $\alpha$.  The denaturated state consists of an
astronomical numbers of microscopic conformations, whereas the native
state is unique. The fact that the unique native state anyway is
reached fairly fast from this large number of unfolded configurations
is known as the Levinthal paradox. Thus, the folding proteins must have
some property encoded which lower configurational entropy
systematically as folding proceeds.  However, such guided folding has
to be merged with the fact that the folding process exhibits
effectively two-state behaviour, i.e.\ has few or no detectable
intermediates \cite{ngnssf97}, respectively $\alpha \sim 4$
\cite{p97}.

Minimalization of folding time and the van't Hoff parameter $\alpha$
poses seemingly contradictory goals. This one might expect
because absence of significant intermediates 
restricts the possibility of guided folding
between the denatured states and the native state.
As an example the following Hamiltonian has $\alpha=4$,
i.e., a maximally sharp transition \cite{bhhsj00},
\begin{equation}
\label{eq:audun}
{\cal H} \; =\; -N \; \phi_1 \phi_2 \cdots \phi_N\;.
\end{equation}
However, it has absolutely no guiding at all:
When in the denaturated
state, i.e., when at least one $\phi_i=0$, there is no energy gain 
in turning this to $\phi_i=1$, unless all other variables $\phi_j$ already
are equal to one.  There is no energy gain unless such a protein moves into
the native state.  Searching for the native state in this system will
grow exponentially with system size $N$.  
In general, it may seem that the more ``two state" the transition is, 
the less guiding can be involved, and
hence, the longer the folding time will be.

In order to discuss folding times,
we need an operational definition of folding time.
We base our definition on the one-step Monte Carlo method \cite{b87}.  
The Monte Carlo method is a way to generate a biased random walk through 
configuration
space such that the relative frequence of occupancy of any configuration
is proportional to the Boltzmann factor.  
Thus Monte Carlo ``time", measuring the number of
steps of the random walk, has strictly speaking nothing to do with ``real" 
time.  However, in a discrete system, such as our model, we believe that 
time defined in this way, essentially brings us as close as possible to
time in a corresponding continuous system.  Hence, we will base our definition
of time in the model by one-step Monte Carlo. 
Hence, we define one unit of time as the
times where all variables have been updated once in average. 
This definition of time is common when dealing with
dynamical questions concerning discrete system \cite{b87}.  

The non-guided Hamiltonian (\ref{eq:audun}) will fold in a 
time proportional to $2^N/N$ when starting from a random initial configuration. 
The fully guided (zipper) Hamiltonian (\ref{eq:singpathphi})
will similarly fold in a time proportional
to $N$. The Baldwin-Rose picture of hierarchical folding \cite{br99a,br99b}
will give a folding time proportional to $\log{N}$. 
These examples, as noted above, seem to suggest
that Hamiltonians with larger $\alpha$ fold faster.  However real 
single-domain proteins have $\alpha$ close to its minimum value 4, 
and are still able to fold in fairly short time. 

We have in Section \ref{realmodel}
suggested a class of hierarchical  
folders, and seen that these typically have $\alpha\sim 4$ that is compatible
with two-state behaviour seen in experiments.

In order to examine the folding behavior, we restrict ourselves initially 
to the case of zero temperature, i.e., when a term has folded it never opens 
again.  The fact that we at all can reach the ground state at zero temperature 
follows from all terms in the Hamiltonian being attractive 
(i.e., negative signs) and that the various terms in the
Hamiltonian guide the system toward the uniquely defined ground state
(where all $\phi_i=1$).
Thus, there are no energy barriers in our idealized model system, and 
folding is therefore only limited by degeneracy/entropy barriers.

The zero temperature limit makes it possible to characterize 
the folding as function
of number of subsequent folding steps, where one step is defined
as an event where at least one term in the Hamiltonian becomes non-zero.
Fig.\ \ref{fig5}a shows the time to fold subsequent steps of the 
Hamiltonian for a system of size $N=100$. 
The ($+$) and ($x$) symbols denote trajectories of two different
Hamiltonians of type (\ref{eq:calhissum}), 
whereas the full line show the ensemble averaged
folding time (over all Hamiltonians and trajectories).
We note that although the trajectories typically fold subsequent levels
of the Hamiltonian, this is not necessarily the case, 
thus opening for alternate pathways. 

Entropy barriers are caused by the necessity to fold several $\phi_i$
terms simultaneously, and the chance 
to do this decreases as the $1/2^k$, where $k$ is the number of terms.
In Fig.\ \ref{fig6}a we display the entropy barrier $dS/dk$ defined as the 
number of simultaneous folding terms
versus the folding step $k$ for the $N=100$ system.
The histogram is $dS/dk$ for one particular Hamiltonian,
whereas the dashed line shows the barriers for the ensemble averaged values.
Fig.\ \ref{fig6}b displays a data collapse of $dS/dk$ for three different
system sizes, demonstrating that both the position and the maximal
barrier scale with system size approximately as $\sqrt{N}$,
which implies that the folding time grows with systems size as $2^{\sqrt{N}}$.
An argument for the obtained scaling is that
after $k$ folding steps, then as long as $k$ is small, 
one typically has folded $k^2/2$ terms. When $k^2/2$ becomes comparable to $N$,
overlap between subsequent steps becomes significant, and subsequent
folding involves fewer new folding variables and therefore 
folding becomes easier. 

In order to estimate whether these numbers are consistent with a fine grained
behavior of real proteins
then the degeneracy of each degree of freedom 
should be of the order of 6, corresponding to the orientational possibilities
of a single amino acid on a random peptide chain.  Thus exchanging $2\to 6$ in
the above analysis, and noting that individual amino acids change conformations
on a timescale of the order of nanoseconds \cite{c93}, 
we very roughly estimate a folding time of
$6^{\sqrt{N}} \times 10^{-9} {\rm s} \sim 10^{-1}$ s for $N=100$
which is of the order found for proteins.

Fig.\ \ref{fig7}a examines 
the zero temperature energy as function of folding step for a 
specific $N=40$ system,
where we associate one unit of energy to each contact term in the Hamiltonian.
Thus the coefficients $a^{(i)}=1$, implying that
all Hamiltonians of type (\ref{eq:calhissum}) have equilibrium transition 
temperature $T_c=1/\log2$. 
The figure shows that the first $k\sim 12$ steps is associated with a slow
linear decrease in energy, hereafter a few subsequent steps 
is responsible for $\sim 65$\% of the total energy gain due to folding.
Combining this with the fact that nearly all entropy 
reduction has to take place for $k<12$ (see Fig.\ \ref{fig6}), 
we conclude that entropy is reduced before energy is gained.

We now examine the effect of finite temperature where
the increased importance of entropy 
converts entropy barriers into free energy barriers. 
This makes the initial folding steps thermodynamically 
disfavoured at temperatures below the equilibrium
melting temperature $T_c$. In Fig.\ \ref{fig7}b this is examined quantitatively,
where free energy is shown versus an effective order parameter $u^*$. 
defined as the number of $\varphi$ variable which is equal to one and belongs
to a term in the Hamiltonian which contributes a finite amount (i.e. $E=1$) 
to the energy. Thus $u^*$ is linearly decreasing with 
the residual entropy of the system.
The plot show the free energy profile for three temperatures,
the $T=0$ case, the temperature where
folding becomes dynamically suppressed, 
$T \;=\; T^* \;=\; 0.27 T_c$ (for $N=40$), 
and the equilibrium denaturation temperature ($T_c = 1/\log2$) 
where the ground state just balances the denatured state. 
We conclude that in our simplified scenario, the sharpness of 
the transition at $T_c$ arises because of an entropy barrier
that will make folding much slower in some limited interval below the
equilibrium melting temperature. With the parameters
of the figure one first obtains the fairly fast folding examined in Fig.\ 
\ref{fig5} when the temperature becomes below $T^*\sim 0.27 T_c$.

We stress that one in principle could include energy barriers into
our formalism by adding positive terms to the hamiltonian
(i.e. $-a_{ijk...}^{(n)}>0$ in Eq.\ (\ref{eq:calhissum})).
Such terms could arise either if the denatured state has binding 
contacts which are not in the
native state or if the transition state have non native contacts. 
Here we focussed on the possibility for a transition
that is purely limited by entropy barriers, thereby defining a transition
state which has little entropy and large enthalpy difference to the ground state,
a feature observed in for $\alpha-$spectrin SH3 by \cite{mvbwmfs99}.
As a consequence we predict a temperature dependence that is opposite
to recent measurement on protein folding kinetics, where
typically the folding time decrease with temperature $T$ for $ T< \sim 40^o$C
\cite{mvbwmfs99,sb97,otsf98}.
Energy barriers therefore plays a role at least at these lower temperatures.
Future measurement of folding
time behaviour close to the denaturation temperature could 
teach us whether the two-state behaviour found for folding of small proteins
is due primarily to enthalpic or entropic barriers.
%------------------------------------------------------------------------------
\section{Conclusion}
\label{conc}
%------------------------------------------------------------------------------
In this paper we have explored the 
idea that entropy correlations and
barriers may appear at any step during a sequential folding process. 
We have seen that this is natural to expect as consequence
of a hierarchical folding process where different branches of the 
hierarchical folding structure are correlated.
We have showed that this folding scheme can result in two-state thermodynamics
even though the energy gain is microscopically small at each folding step.
We have seen that this two state thermodynamics can be merged with
a fairly fast folding process, with time increasing with $N$ as
$2^{\sqrt{N}}$ instead of 
$2^{N}$ as one naively should expect for two-state folders.
Further we have found the qualitatively same subdivision between initial
and final folding as reported in \cite{bhhsj00}, without the need to resorting
to a special treatment of any terms.  
This subdivision into early entropically disfavoured 
guiding and last step easy match 
predicts a folding time that increases with 
temperature when denaturation is approached.

We thank A.\ Bakk and J.\ S.\ H{\o}ye for many fruitful discussions.  P.\ G.\
D.\ thanks the NFR for financial support.  P.\ G.\ D.\ and A.\ H.\ thank
Nordita and the Niels Bohr Institute for friendly hospitality and
support.
%------------------------------------------------------------------------------
\appendix
\section{Calculating the average partition function}
\label{ap:average}
%------------------------------------------------------------------------------
We can make $N!/m!$ different products of $m$ $\phi_i$. Thus, there is in
total 
\begin{equation}
\label{eq:misfractn}
M=\frac{N!}{(N-1)!} \frac{N!}{(N-2)!}\dots\frac{N!}{(N-N)!}
\end{equation}
different Hamiltonians (note, however, that most of them are equal up
to permutations of $\phi_i$).

We may calculate the mean value of the partition function
\begin{eqnarray}
\langle Z \rangle &=& \frac{1}{M}\sum_{ \{a \}}\sum_{ \{ \phi \}}{\rm
e}^{-{\cal H}/T} \nonumber \\
  &=& 1+\frac{1}{M}\sum_{ \{ \phi \}}\sum_i {\rm
  e}^{\phi_i/T}\sum_{i\ne j} {\rm e}^{\phi_i
\phi_j/T}\dots\sum_{i\ne j \ne k\dots}
{\rm e}^{\phi_i \phi_j \phi_k \dots/T}
\end{eqnarray}
The Hamiltonian is invariant to all permutations of
$\phi$, and this suggests the new variable $s=\sum_{i=1}^N\phi_i$.
We then find 
\begin{equation}
\sum_i {\rm
e}^{\phi_i/T}=\frac{s!}{(s-1)!}{\rm}^{1/T}+\frac{N!}{(N-1)!}-
\frac{s!}{(s-1)!}\;,
\end{equation}
and
\begin{equation}
\sum_{i\ne j} {\rm e}^{\phi_i
\phi_j/T}=\frac{s!}{(s-2)!}{\rm}^{1/T}+\frac{N!}{(N-2)!}-\frac{s!}{(s-2)!}\;,
\end{equation}
and so on. This gives us an expression for the mean partition function
\begin{equation} 
\label{eq:meanz}
\langle Z \rangle = 1+\frac{1}{M}\sum_{s=1}^N \left(
\begin{array}{c} N\\s \end{array}\right)\prod_{k=1}^{s}\left[
\frac{s!}{(s-k)!}{\rm
e}^{1/T}+\frac{N!}{(N-k)!}-\frac{s!}{(s-k)!}\right]
\prod_{k=s+1}^{N}\frac{N!}{(N-k)!}\;.
\end{equation}
This expression can be further simplified to
\begin{equation} 
\label{eq:meanz1}
\langle Z \rangle = 1+\sum_{s=1}^N \left(
\begin{array}{c} N\\s
\end{array}\right)\prod_{k=1}^{s}\left[1+\frac{s!(N-k)!}{N!(s-k)!}({\rm
e}^{1/T}-1)\right]\;.
\end{equation}
%------------------------------------------------------------------------------
\section{Fluctuations of Partition Function}
\label{ap:selfaverage}
%------------------------------------------------------------------------------
In this Appendix, we find the expression for $\langle Z^2\rangle$
which was used to show that the relative fluctuations $(\langle
Z^2\rangle -\langle Z\rangle^2) /\langle Z\rangle^2$ goes to zero
as $N\to\infty$, as is seen in Fig.\ \ref{fig2}.  We have that
\begin{eqnarray}
\label{eq:calz2}
\langle Z^2 \rangle &=& \frac{1}{M}\sum_{ \{a \}}\sum_{ \{
\phi \} }\sum_{ \{ \phi' \}}{\rm e}^{-H(\phi)/T - H(\phi')/T} 
\nonumber \\ &= & \sum_{ \{ \phi \} }\sum_{ \{
\phi' \}} G(\phi,\phi')\;,
\end{eqnarray}
where 
\begin{equation}
\label{eq:gvar}
 G(\phi,\phi') = \sum_{ \{ \phi \}}\sum_{ \{ \phi'
  \}}\sum_i {\rm e}^{(\phi_i+\phi'_i)/T}\sum_{i\ne j} {\rm
  e}^{(\phi_i \phi_j+\phi'_i \phi'_j)/T}\dots\sum_{i\ne j
  \ne k\dots} {\rm e}^{(\phi_i \phi_j \phi_k \dots+\phi'_i
  \phi'_j \phi'_k \dots)/T}\;.
\end{equation}
The sum over $\phi$ and $\phi'$ is simplified by noting that
the function $G(\phi,\phi')$ is symmetric to all permutations of
pairs $(\phi_i, \phi'_i )$. This means that $G$ can be expressed
as
\begin{equation}
\label{eq:gphi}
G(\phi,\phi')=G(s,s',p)\;,
\end{equation}
with $ s = \sum_{i=1}^N\phi_i $, $s'= \sum_{i=1}^N\phi'_i $ and
$ p = \sum_{i=1}^N \phi_i \phi'_i$. The sum over all $\phi_i$
and $\phi'_i$ is thus reduced to a sum over $s$, $s'$ and $p$.
\begin{eqnarray} 
\label{eq:z2}
\langle Z^2 \rangle &=& \frac{1}{M} 
\sum_{s=0}^N\sum_{p=p_{\rm min}}^s \left( \begin{array}{c} N\\p
\end{array}\right)\left( \begin{array}{c} N-p\\s-p
\end{array}\right)\left( \begin{array}{c} N-s\\s-p \end{array}\right)
G(s,s,p) \nonumber \\ &+&\frac{1}{M}
2\sum_{s=1}^N\sum_{s'=0}^{s-1}\sum^{s'}_{p=p_{\rm min}} \left(
\begin{array}{c} N\\p
\end{array}\right)\left( \begin{array}{c} N-p\\s-p
\end{array}\right)\left( \begin{array}{c} N-s\\s'-p
\end{array}\right)G(s,s',p)\;,
\end{eqnarray} 
where $p_{\rm min}=(s+s'-N)\Theta(s+s'-N)$. The function $G(s,s',p)$
is a product of $N$ sums. Each sum can be be evaluated. The results is
\begin{equation}
\label{eq:gss}
G(s,s',p)=\prod_{m=1}^N A_m(s,s',p)\;,
\end{equation}
where we have that
\begin{eqnarray}
\label{eq:amss}
A_m(s,s',p) &=& \Theta(p-m)\frac{p!}{(p-m)!}{\rm e}^{ 2/T} \nonumber
\\ &+&\left[ \Theta(s-m)
\frac{s!}{(s-m)!}-\Theta(p-m)\frac{p!}{(p-m)!} \right ] {\rm e}^{1/T}
\nonumber \\ &+&\left[ \Theta(s'-m)
\frac{s'!}{(s'-m)!}-\Theta(p-m)\frac{p!}{(p-m)!} \right ] \nonumber\\
{\rm e}^{1/T} \nonumber \\ &+& \frac{N!}{(N-m)!}+\Theta(p-m)
\frac{p!}{(p-m)!}-\Theta(s-m)\frac{s!}{(s-m)!}-
\Theta(s'-m)\frac{s'!}{(s'-m)!}\;.
\end{eqnarray}
Here $\Theta(n)$ is the step function ($\Theta(n)=1$ for $n\ge 0$).
%------------------------------------------------------------------------------

% --------------------------------------------------------------------
% FIGURE CAPTIONS
% --------------------------------------------------------------------
\begin{figure}
\caption{Folding pathways shown as a directed network between partially
folded structures, which have been numbered.
\label{fig1}
}
\end{figure}

\begin{figure}
\caption{Folding pathway network corresponding to the hierarchical folding
scheme described by Baldwin and Rose.
\label{fig1.5}
}
\end{figure}

\begin{figure}
\caption{Relative fluctuations of partition function 
for different temperatures as a function of system size $N$.  This shows that
the system is self averaging strongly enough, so that $\log\langle Z\rangle
=\langle \log Z\rangle$.
\label{fig2}
}
\end{figure}

\begin{figure}
\caption{Order parameter $u$ for $N=50$, 100, 200 as a function of temperature
$T$.
\label{fig3}
}
\end{figure}

\begin{figure}
\caption{$\alpha$ coefficient for different system sizes $N$.
\label{fig4}
}
\end{figure}

\begin{figure}
\caption{a) Folding trajectories of $N=100$ systems.
We show time since denatured state
as function of folding step $k$. 
If all terms in Hamiltonian are assigned equal energy $a=1$, each
folding step is associated by one unit of energy reduction.
The ($+$) symbols denote three trajectories for one Hamiltonian,
the ($\times$) symbols show three trajectories for another Hamiltonian.
The full line shows the ensemble
averaged folding trajectory. One observes
that full folding is reached after about $3\cdot 10^5$ timesteps,
and that it only require the first $k \sim 20$ terms in the Hamiltonian
to be explicitly folded. 
\label{fig5}
}
\end{figure}

\begin{figure}
\caption{a) Folding barrier as function of folding step 
for the $N=100$ system. The largest folding barrier is caused by 
the maximum number of $\phi$-variables that must be set to one without 
guiding in order to make a term in the Hamiltonian non-zero. 
The ($+$) symbols illustrate a case where this barrier involves folding of
8  $\phi$ variables at folding step $k=9$.
b) Rescaled folding barriers ($(N/100)^{1/2} dS/dk$)
versus rescaled folding step ($(N/100)^{1/2} k$) for three different
system sizes ($N=25$, 100 and 400).
\label{fig6}
}
\end{figure}

\begin{figure}
\caption{a) Example of energy as function of folding step $k$
for a system of $N=40$ folding variables and all terms 
in Hamiltonian assigned a prefactor $a^{(i)}=1$.
Note the distinct difference between early and late folding regime,
contrasting the linear decrease found for the simple zipper model.
b) Free energy versus effective order parameter (defined in Eq.\ 
(\ref{eq:uisvar}))
for three different temperature. Note that the monotonic decrease at low
temperature is replaced with a free-energy barrier 
at intermediate temperatures.
\label{fig7}
}
\end{figure}
% --------------------------------------------------------------------
\begin{table}
\begin{tabular}{|c|ccccc|}
 &$\phi_1$&$\phi_2$&$\phi_3$&$\phi_4$&$\phi_5$\\
\hline
1&1& & & &5\\
2& & & &4&4\\
3&2&2& & &2\\
4& & & & & \\
5&3&3&3&3&3\\
\end{tabular}
\caption{Folding table for the pathway network shown in Fig.\ \ref{fig1}. 
The columns refer to the variables $\phi_i$, while the rows refer to
products of $\phi_i$ variables.  The left-most column refers to the number
of $\phi_i$s in the products. The numbers to the right of the
left-most column refer to the five structures in Fig.\ \ref{fig1}.
\label{table1}}
\end{table}
% --------------------------------------------------------------------
\end{document}